\documentclass[aps,
               prb,
               reprint,
               amsfonts,
               unsortedaddress,
               superscriptaddress,
               amssymb,
               showkeys]{revtex4-1}

\usepackage{amsmath}      
\usepackage{graphicx}     
\usepackage{color}        
\usepackage{hyperref}     

\newcommand{\diffconst}{\ensuremath{D}}
\newcommand{\difffreq}{\ensuremath{\nu_s}}
\newcommand{\evafreq}{\ensuremath{\nu_e}}
\newcommand{\meltrad}{\ensuremath{r_m}}
\newcommand{\teflonr}{Teflon\textsuperscript{\textregistered}}
\newcommand{\teflon}{Teflon}
\newcommand{\expnum}[2]{\ensuremath{#1\mspace{-1mu}\times\mspace{-1mu}10^{#2}}}
\newcommand{\cel}[1]{\ensuremath{#1\,^{\circ}\mathrm{C}}}
\newcommand{\CAU}{Christian-Albrechts-Universit\"at zu Kiel}
\newcommand{\fref}[1]{Fig.\,\ref{#1}}
\newcommand{\eref}[1]{Eq.\,\ref{#1}}
\newcommand{\cond}{\ensuremath{C}}
\newcommand{\rate}[1]{\ensuremath{{R}_{#1}}}
\newcommand{\Rate}{\ensuremath{{R}}}

\newcommand{\ITAP}{Institut f\"ur Theoretische Physik und Astrophysik, \CAU, Leibnizstra{\ss}e 15, D-24098 Kiel, Germany}

\newcommand{\MAWI}{Institut f\"ur Materialwissenschaft, Lehrstuhl f\"ur Materialverbunde, \CAU, Kaiserstra{\ss}e 2, D-24143 Kiel, Germany}

\newcommand{\GREVE}{Now: NXP Semiconductors, Hamburg, Germany}

\newcommand{\MAILFAU}{ff@tf.uni-kiel.de}
\newcommand{\MAILBON}{bonitz@physik.uni-kiel.de}
\newcommand{\INST}[1]{\affiliation{#1}}

\begin{document}

\title{Formation of magnetic nanocolumns during vapor phase deposition of a metal-polymer nanocomposite:
       experiments and kinetic Monte Carlo simulations}

\author{L. Rosenthal}
\INST{\ITAP}

\author{H. Greve}
\INST{\MAWI}
\INST{\GREVE}

\author{V. Zaporojtchenko}
\thanks{deceased}
\noaffiliation

\author{T. Strunskus}
\INST{\MAWI}

\author{F. Faupel}
\email[mail to: ]{\MAILFAU}
\INST{\MAWI}

\author{M. Bonitz}
\email[mail to: ]{\MAILBON}
\INST{\ITAP}

\date{\today}

\begin{abstract}
Metal-polymer nanocomposites have been investigated extensively during the last years due to their interesting functional applications. They are often
produced by vapor phase deposition which generally leads to the self-organized formation of spherical metallic nanoparticles in the organic matrix,
while nanocolumns are only obtained under very specific conditions. Experiments\cite{Grev+06} have shown that co-evaporation of the metallic and
organic components in a simple single-step process can give rise to the formation of ultrahigh-density Fe-Ni-Co nanocolumnar structures embedded
in a fluoropolymer matrix. Here we present a kinetic Monte Carlo approach which is based on an new model involving the depression of the melting
point on the nanoscale and a critical nanoparticle size required for solidification. In addition we present new experimental results down to a
deposition temperature of \cel{-70} and also report the magnetic properties. The simulations provide a detailed understanding of the transition from
spherical cluster growth to formation of elongated structures and are in quantitative agreement with the experiments.
\end{abstract}
\maketitle

\section{\label{sec:intro}Introduction}
Nanocomposites containing metallic nanoparticles in a dielectric polymer matrix have very interesting functional applications ranging from plasmonics
\cite{caseri_nanocomposites_2000,elbahri_omnidirectional_2011} and high-frequency magnetic materials\cite{greve_nanostructured_2006} to antibacterial
coating\cite{zaporojtchenko_physico-chemical_2006,hahn_therapeutic_2010} (for recent reviews see\cite{faupel_metal-polymer_2010,carotenuto_metal-polymer_2005,
hanemann_polymer-nanoparticle_2010}). Such metal-polymer nanocomposites are often produced by vapor-phase co-deposition of the metallic and organic
components\cite{heilmann_xps_1997,biederman_rf_2000,schurmann_controlled_2005}, since this allows tailoring of the nanoparticle filling factor and
other parameters, and even allows incorporation of alloy nanoparticles with well-defined composition\cite{beyene_vapor_2012}.
During co-deposition, metallic nanoparticles form by self-organization due to the much higher cohesive energy of the metal compared to the organic
component and the low metal organic interaction energy (except for very reactive metals).
One can assume that the self-organization mechanism during the deposition of polymer-based nanocomposites is analogous to metal cluster formation
on a polymer surface\cite{Faupel_conf_1999}. When energetic metal atoms impinge on the polymer surface they undergo various processes including random
walk on the surface, diffusion into the bulk, and desorption\cite{Zapo_conf_2000,zaporojtchenko_metal_polymer_2000}. Within their diffusion distance,
metal atoms may encounter each other or may be captured by a surface defect. This leads to aggregation and formation of stable metal clusters which
are embedded into in the polymer matrix upon growth of the nanocomposite film. The metal filling factor depends on the condensation coefficient of
metal atoms on a given polymer surface\cite{thran_condensation_1999} as well as on the metal-polymer deposition ratio\cite{takele_plasmonic_2006}.
In these terms, the volume fractions of metallic nanoparticles in the composite films can be easily controlled through the ratio of the deposition
rates of metal \rate{m} and polymer \rate{p} components.

Generally, the nanoparticles obtained upon vapor phase co-deposition have spherical shape (as long as the filling factor is small enough to prevent
nanoparticle coalescence). This is expected form the minimization of the surface energy and the above mentioned formation process. However, under
specific conditions, involving a very high deposition rate ratio of the metallic over the organic components and a very low metal condensation
coefficient, formation of elongated Fe-Ni-Co\cite{Grev+06} and Au\cite{biswas_2006} nanocolumns has been reported in a Teflon AF matrix. While
the formation process of the spherical nanoparticles is well understood and has also been modeled by kinetic Monte Carlo simulations\cite{silverman_1991,
thran_1997,faupel_diffusion_1998,rosenthal_diffusion_2011,bonitz_towards_2012} only a crude qualitative model was suggested for nanocolumn
formation\cite{Grev+06}. According to this model, a very low condensation coefficient is crucial. Thus the metal atoms arriving on the growing
nanocomposite film from the gas phase will stick whenever they encounter a growing metallic nanoparticle but will have a very large surface
diffusivity and a high thermal desorption probability if they impinge on the organic matrix, due to the very low metal-organic interaction energy.
On the organic surface, there is a competition between thermal desorption and diffusion to and trapping at a nanoparticle. In these terms, it was
proposed that at a critical metal vs organic deposition rate ratio, metal nanoparticles can grow faster perpendicular to the surface, due to direct
impingement of newly arriving metal atoms from the gas phase, then they are embedded by the growing organic matrix. However, kinetic Monte Carlo
simulations based on this notion failed to reproduce the experimental results even qualitatively.

Here we report new experimental results on the formation of Fe-Ni-Co nanocolumns in a Teflon AF matrix via co-evaporation, which extend the
temperature range of the previous experiments down to \cel{-70} and include the magnetic characterization of the highly anisotropic properties.
In addition, we propose a new model for nanocolumn formation and we present new kinetic Monte Carlo simulations which are able to explain the
experimental results even quantitatively. A key new aspect of the present model is solidification of the nanoparticles at a critical size, which
drastically slows down the kinetics for the establishment of the spherical equilibrium shape.

\section{\label{sec:experiments}Experiments}
\begin{figure}[t!]
  \includegraphics{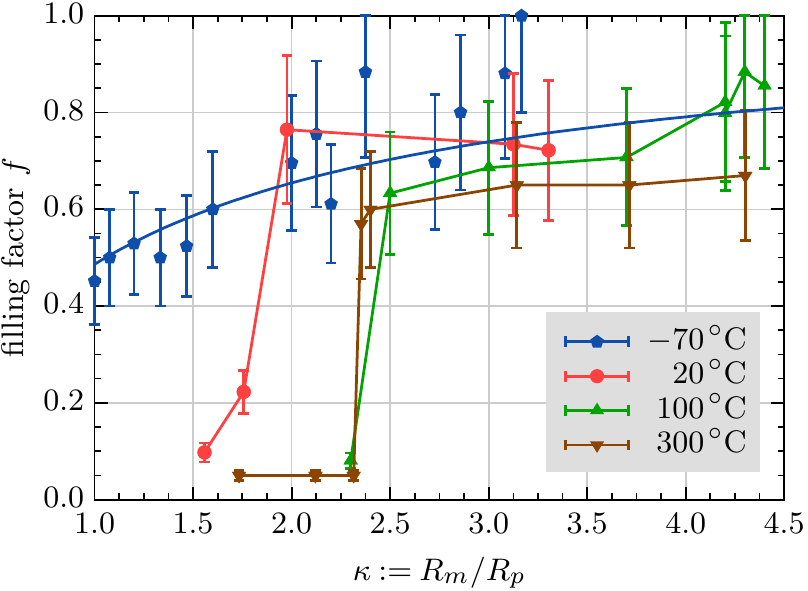}
  \caption{\label{fig:fill_fac_exp} (Color online) Experimental results of the volume filling factor as a function of the ratio $\kappa$ of deposition rates
           \Rate~of Fe-Ni-Co and Teflon for different temperatures.}    
\end{figure}

The nancomposite films of thicknesses 100\,nm to 200\,nm were produced by co-evaporation of the organic and metallic components on Si
wafers using a homemade high vacuum deposition chamber\cite{biswas_2003,Grev+06}. \teflonr AF (granulates, Dupont) and Fe-Ni-Co (99.99\% pure
1\,nm diameter wires, Good Fellow Industries, U.K.) were used as starting materials. For preparation of samples for transmission electron
microscopy (TEM) and magnetic characterization, polymer foils (Upilex-S\textsuperscript{\textregistered}) were used as substrates. Polymers
generally do not lend themselves for evaporation because they decompose upon heating, however, for some polymers such as Teflon AF, the monomer
structure is preserved upon thermal breaking of the covalent bonds along the backbone chain of the polymer, and a Teflon AF film can deposited
which differs from the starting material mainly by its much lower molecular weight. The molecular weight reduction is not critical for functional
applications. Deposition rates of $0.15-0.3$\,nm/min and $0.6-1$\,nm/min were typically used for \teflon AF and Fe-Ni-Co, respectively.
The metallic volume filling factor of the nanocomposites was determined by energy dispersive X-ray spectroscopy (EDX) as described in
Ref\cite{schurmann_controlled_2005}. The experimental error is $\pm$20\%. The magnetic measurements were carried out with a LakeShore 7300
vibrating sample magnetometer (VSM). For further experimental details see\cite{Grev+06,biswas_2003,greve_diss}.
\fref{fig:fill_fac_exp} shows the metal volume filling factor f as function of the deposition rate ration $\kappa = \rate{m}/\rate{p}$ of the
metallic vs the polymer components for deposition at different substrate temperatures. At the highest temperatures, one notes a sharp increase
in the filling factor above a critical $\kappa$ value. The increase shifts to lower $\kappa$ values and is more smeared out for lower substrate
temperatures, which is most pronounced at \cel{-70}. At this temperature the data were fitted to the function
\begin{equation}\label{eq:fit_func}
  f=\frac{\kappa\cond}{\kappa\cond+1}
\end{equation}
Here the fitting parameter \cond~is the metal condensation coefficient. Equation \ref{eq:fit_func} follows immediately by expressing f in terms of the effective
deposition rates which are multiplied by the condensation coefficients, taking into account metal desorption, and assuming complete condensation
for the organic component.
\begin{figure}
  \includegraphics[width=4.5cm]{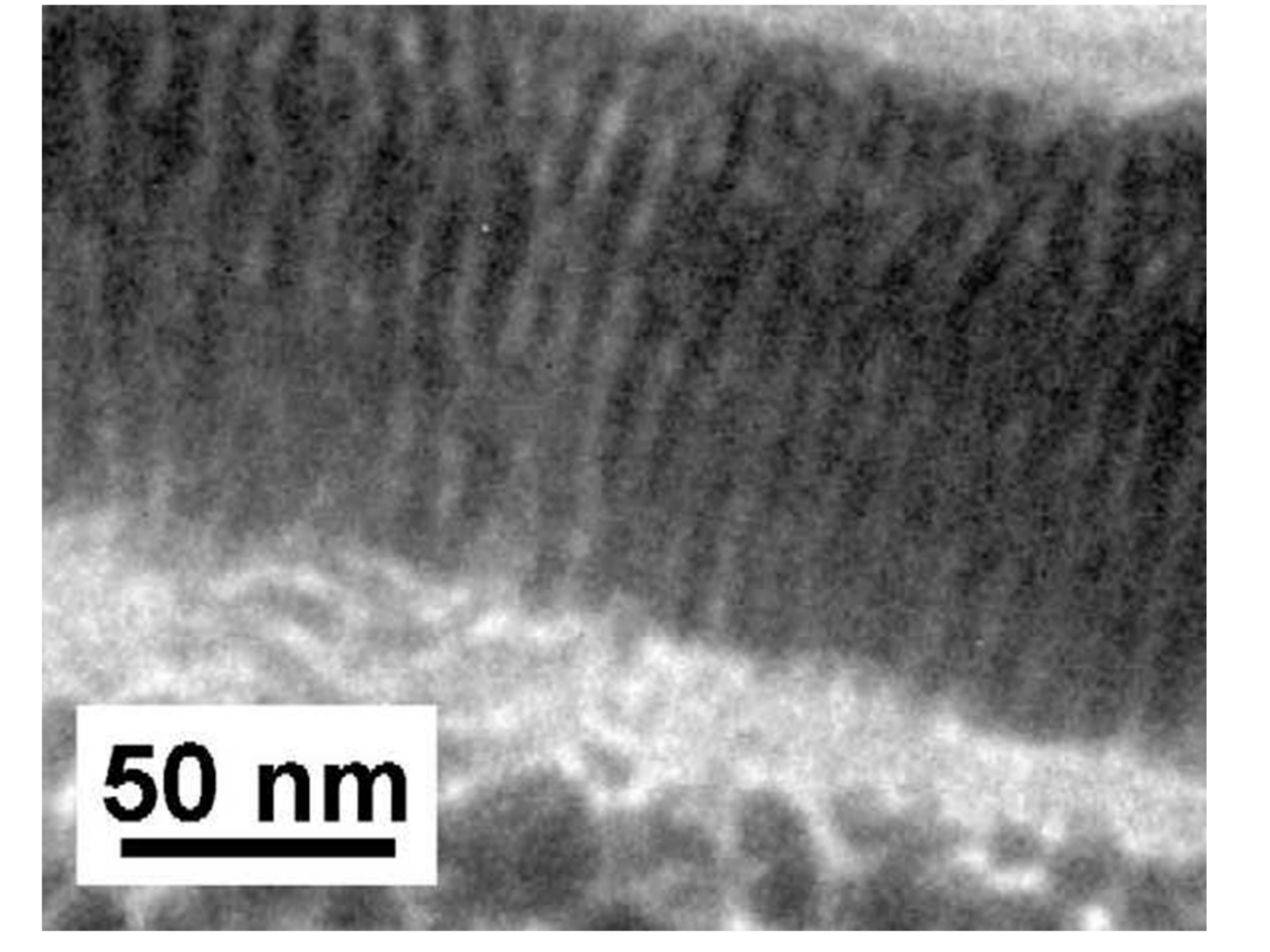}\hspace*{0.15cm} 
  \includegraphics[width=3.285cm]{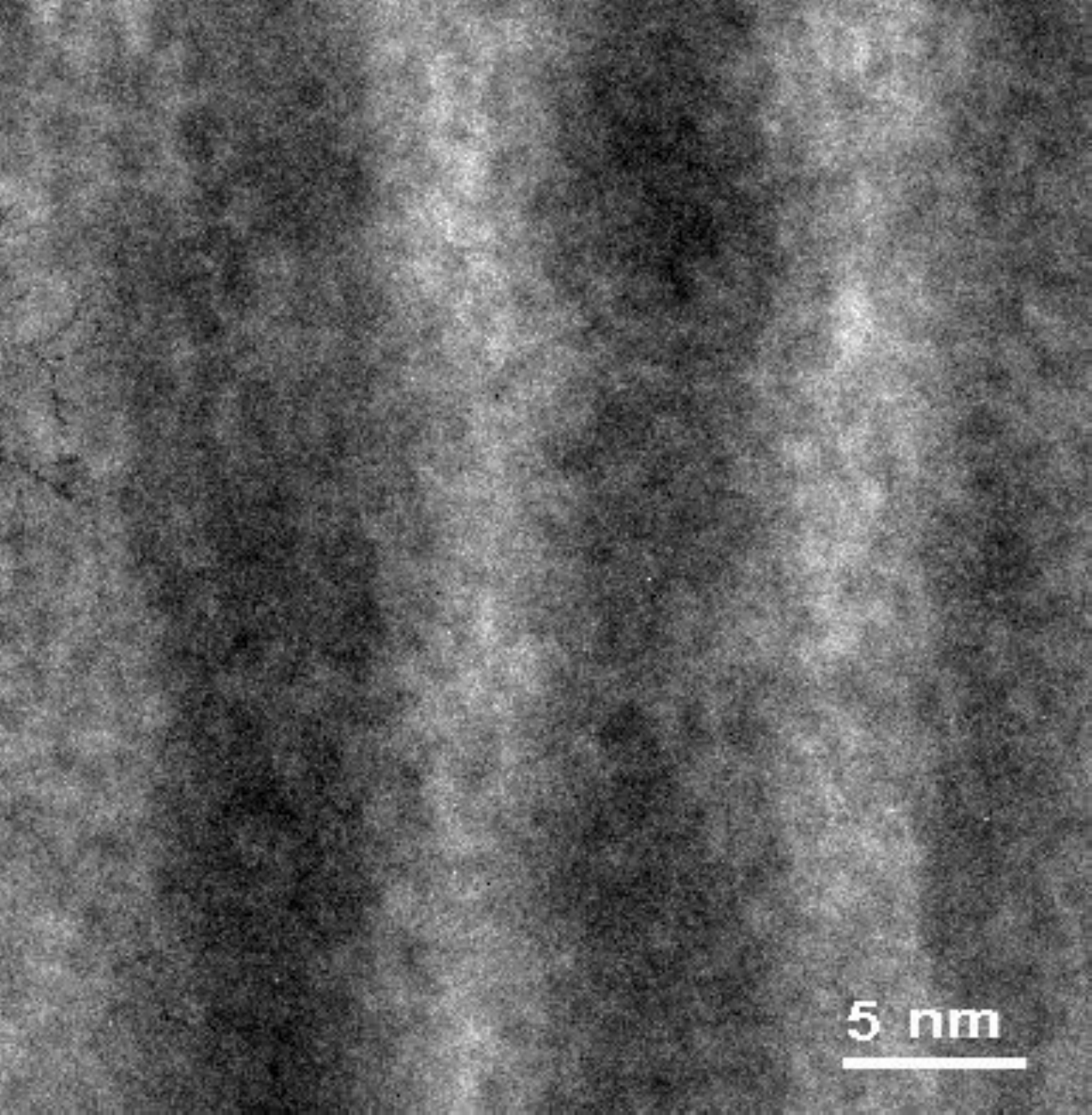}               
  \caption{\label{fig:tem_columns}(a) Cross-sectional TEM image showing the formation of Fe-Ni-Co nanocolumns in Teflon AF on top of a layer of Ag
           clusters in Teflon AF isolated by 20\,nm of the same matrix material. (b) Cross-sectional higher magnified TEM image of
           self-organized nanocolumns of Fe-Ni-Co in Teflon AF.}
\end{figure}
The fit yields $C = 0.94 \pm 0.08$ in good agreement with the expectation that the condensation coefficient approaches
unity at low temperatures\cite{zaporojtchenko_determination_2000}. It has to be pointed out, however, that the condensation coefficient depends on
the metal coverage at the surface of the growing composite and hence on $\kappa$ because metal atoms stick with a probability of unity if they
directly impinge on a metal nanoparticle or if they reach a metal nanoparticle via surface diffusion. Therefore, \eref{eq:fit_func} is not applicable
at higher temperatures, where the condensation coefficient on Teflon AF is expected to be very low\cite{zaporojtchenko_determination_2000,
thran_condensation_1999}. Even the value of $\cond = 0.94$ obtained at \cel{-70} probably overestimates the condensation coefficient for the pure polymer.

\begin{figure}[t!]
  \includegraphics[width=5.0cm]{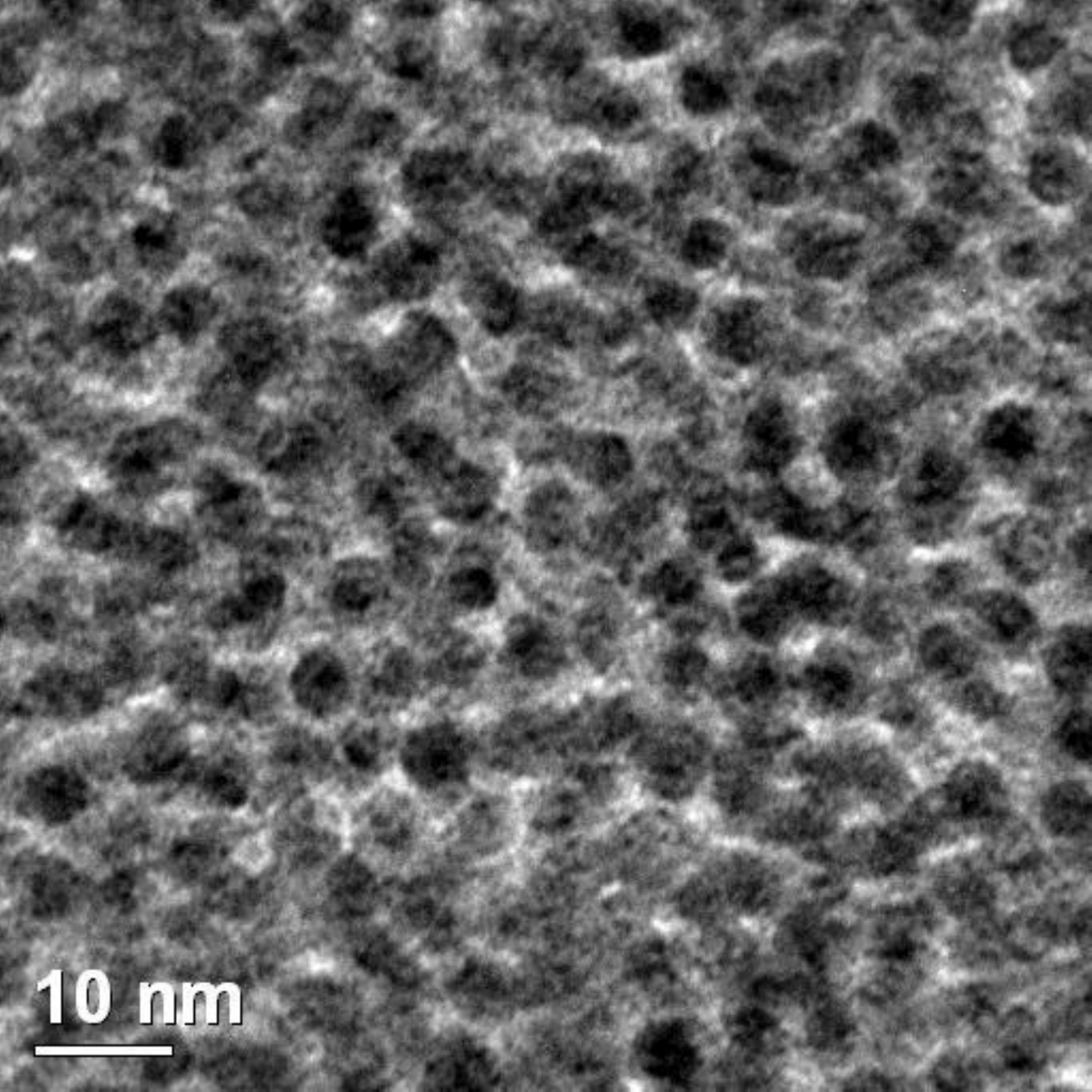}
  \caption{\label{fig:cluster_top_view_exp}Top-view TEM micrograph showing Fe-Ni-Co nanocolums in a Teflon AF matrix. The nanocomposite was deposited
           at a low substrate temperature of \cel{-70} on an electron transparent TEM grid which was covered with a Teflon AF layer prior to the
           nanocomposite deposition to exclude substrate effects\cite{greve_diss}.}
\end{figure}

The microstructure was investigated by means of transmission electron microscopy. Representative TEM micrographs are displayed in
Figs.\,\ref{fig:tem_columns} and \ref{fig:cluster_top_view_exp} (see Ref.\cite{greve_diss} for further details). \fref{fig:tem_columns} shows cross-sectional
images of a nanocomposite prepared at \cel{160}. (As described in Ref.\cite{greve_diss}, the nanocomposite film was grown on top of an evaporated
Teflon AF film containing spherical Ag nanoparticles. In order to exclude any influence of the Ag particles on the growth of the magnetic
nanocolumns, a Teflon AF separation layer of 20\,nm was evaporated on top of the Ag-Teflon AF nanocomposite film before the of the Fe-Ni-Co-Teflon
AF nanocomposite was deposited.)
It is obvious from \fref{fig:tem_columns} that the Fe-Ni-Co nanoparticles have grown as nanocolumns with a diameter of about $7-8$\,nm and a length
extending through the whole film, resulting in an aspect ratio well above 10. The deposition was performed under conditions of normal incidence, and
the orientation of the nanocolumns is perpendicular to the substrate. Experiments were also carried out with normal incidence of the organic component
and with an incident angle of $55^\circ$ with respect to the substrate for the metallic component. Under these conditions, the resulting nanocolumns
were inclined with an angle of $70-75^\circ$ with respect to the substrate, indicating that the growth direction can be controlled via the angle of
incidence.
\fref{fig:cluster_top_view_exp} shows a top-view TEM micrograph of a nanocomposite film deposited at \cel{-70}. The film has a thickness of
30\,nm to ensure electron transparency. Evidence for the nanocolumnar shape of the metallic particles is provided from the fact that no
overlap of particles is seen. For spherical particles of about 5 nm diameter, the electron beam would have a high probability to penetrate
through more than one particle, giving the impression of coalesced particles as always seen in top-view images under the present conditions
for spherical particles\cite{faupel_metal-polymer_2010,takele_tuning_2008}.

\begin{figure}[t!]
  \includegraphics{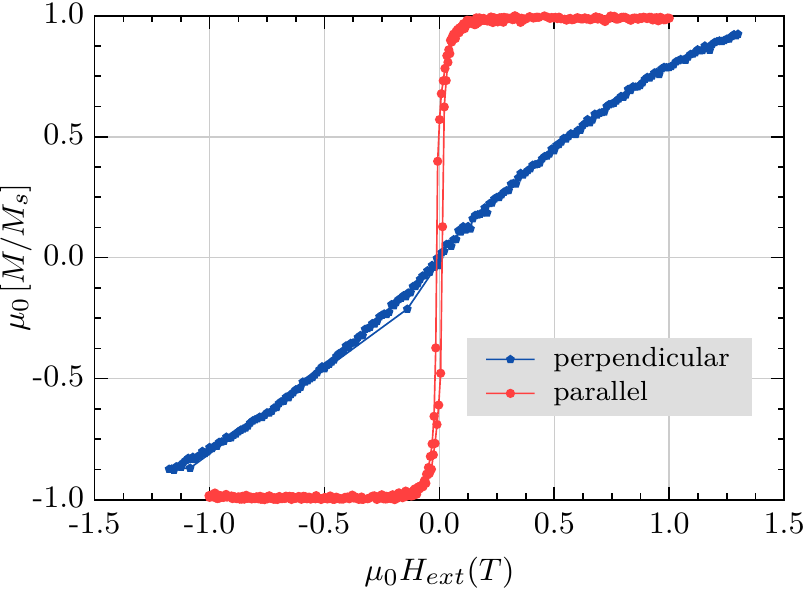}
  \caption{\label{fig:hysteris}Hysteresis curves for a Fe-Ni-Co-Teflon AF nanocomposite film co-evaporated at \cel{300} substrate temperature.
           One notes a strong magnetic anisotropy with the easy axis of magnetization parallel to the film plane.}
\end{figure}

\begin{figure}[b!]
  \includegraphics[width=3.0cm]{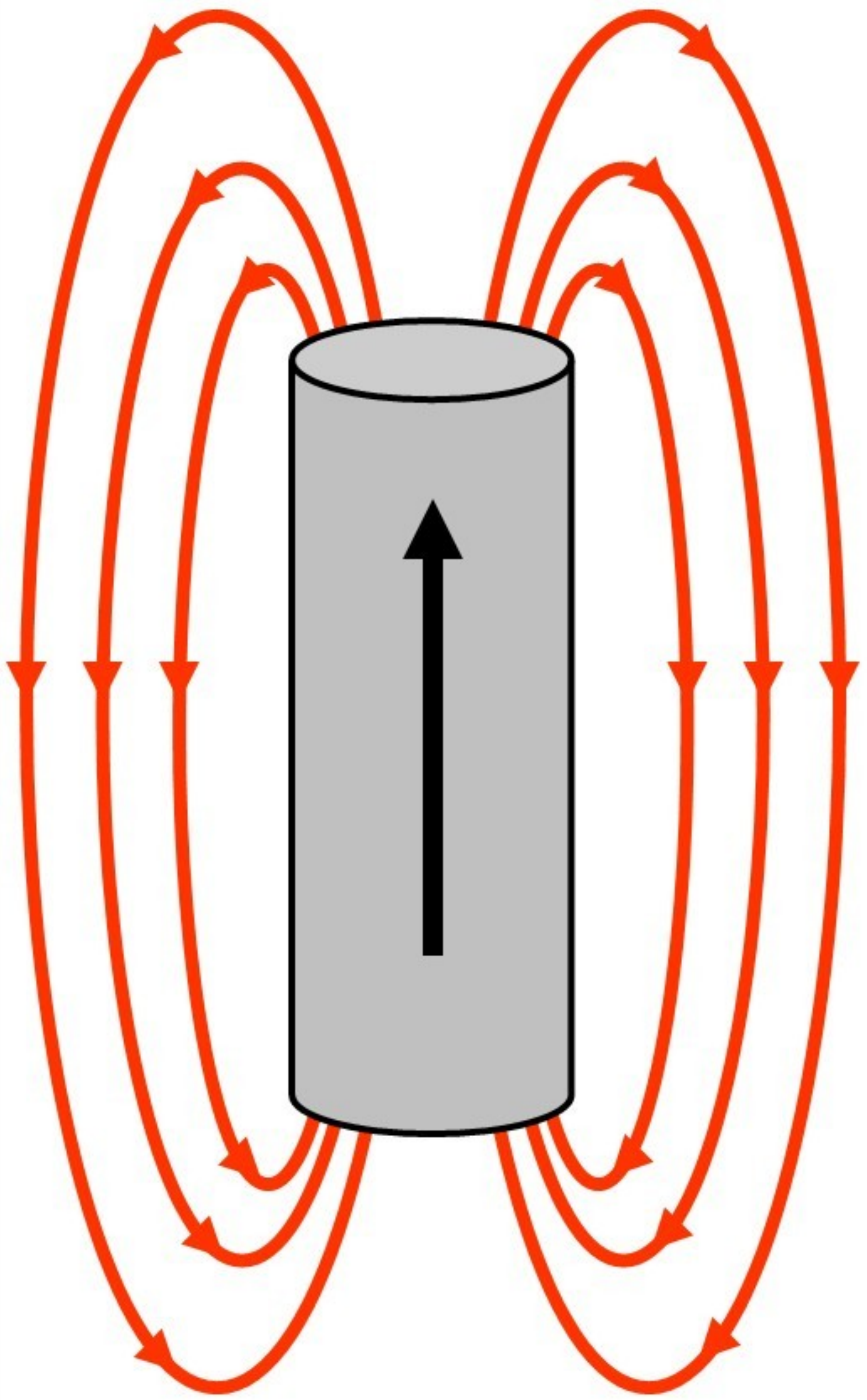}\hspace*{1.0cm}
  \includegraphics[width=4.0cm]{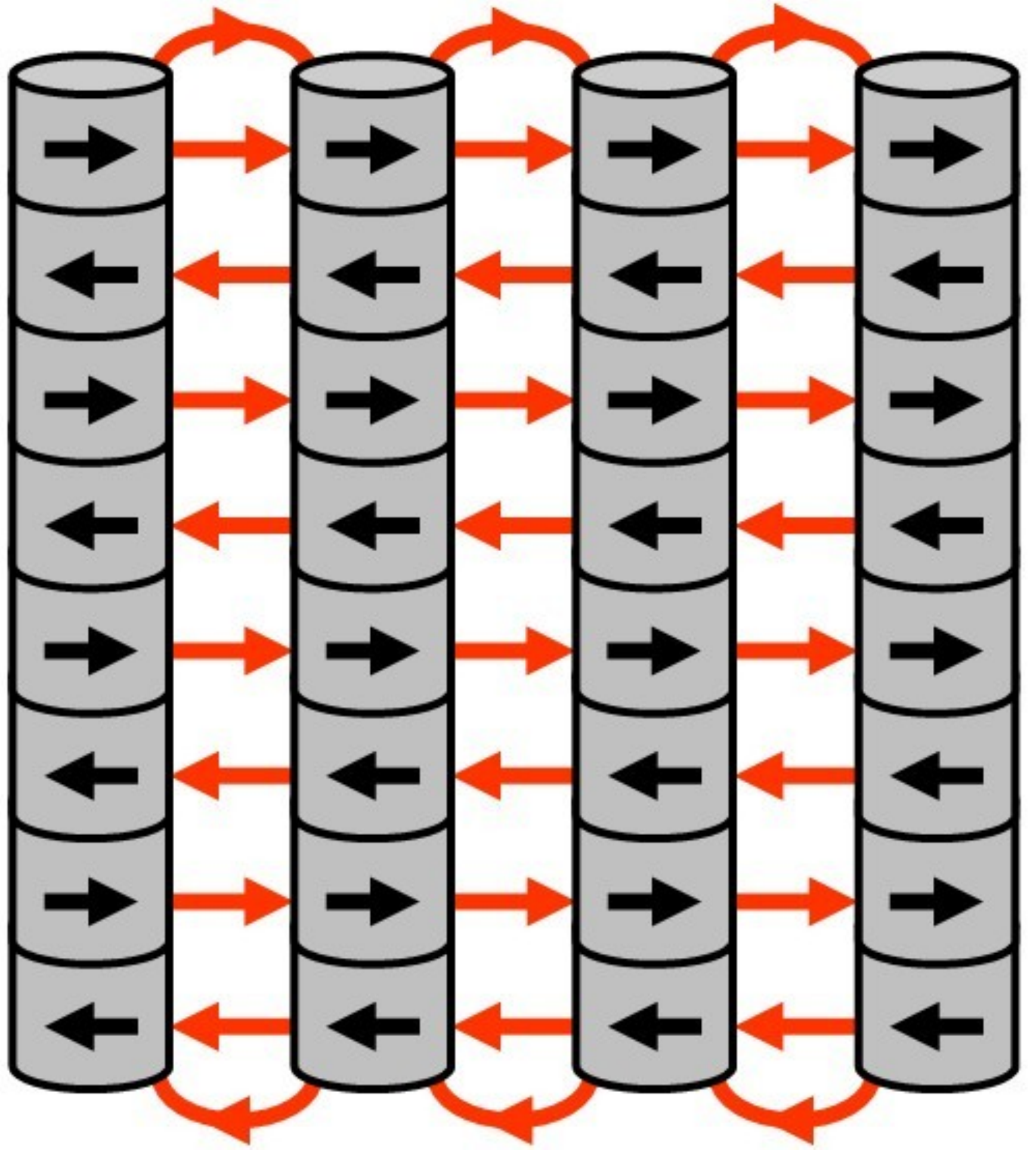}
  \caption{\label{fig:magnetization_sketch}Left: Sketch of the magnetization in an isolated long magnetic nanocolumn, where the easy axis of magnetization
           is parallel to the column due to shape anisotropy. Right: In an array of nanocolumns with strong dipole-dipole coupling, the easy axis is
           oriented perpendicular to the nanocolumns, and magnetic domain walls are incorporated.}
\end{figure}

The highly isotropic nanocolumnar structure is also reflected in highly anisotropic magnetic properties. Hysteresis curves for a sample deposited
at \cel{300} are shown in \fref{fig:hysteris}. One notes a completely different behavior for measurements parallel and perpendicular to the film plane.
The very soft magnetic behavior parallel to the film is a clear signature of magnetization reversal by domain wall movement, whereas saturation
in the perpendicular direction requires very high fields indicating that magnetization reversal is only possible without domain wall
movement\cite{cullity_introduction_2008}. Apparently, the magnetization of the nanocolumns is different from the case of a long isolated
column (\fref{fig:magnetization_sketch}\,(a)) where shape anisotropy always leads to an orientation of the easy axis parallel the column, and reveals the presence of
domain walls perpendicular to the long column axis (\fref{fig:magnetization_sketch}\,(b)).

The observed magnetization behavior can be explained in terms of a competition of demagnetizing fields and dipole–dipole fields. It has been
shown, e.g., for arrays of much larger Co nanowires electrodeposited in anodic alumina, that the magnetization can be tuned parallel or perpendicular
to the nanowires by changing their length\cite{strijkers_structure_1999}. In the present case, the dipole-dipole interaction dominates the behavior
due to the small nanocolumn separation of only a few nanometers.


\section{Kinetic Monte Carlo Simulations}\label{sec:KMC_description}
The Kinetic Monte Carlo simulations described below are based on previous simulations of diffusion and growth of metal clusters in a
polymer substrate\cite{silverman_1991,thran_1997,faupel_diffusion_1998,rosenthal_diffusion_2011,bonitz_towards_2012}. The main idea behind these
simulations is to condense the complex processes of cluster growth and nanocolumn formation during polymer metallization into a simplified
selfconsistent continous-space simulation scheme. To get a practical simulation
algorithm we applied the following assumptions: The atomic and chemical structure of the polymer substrate is essentially neglected. Instead the
influence of the substrate is reflected in averaged cluster mobilities\,(expressed in rate constants) and diffusion jump lengths of clusters.
The polymer is assumed to be a continuum with periodic boundary conditions in the \textit{x}- and \textit{y}-direction\,(parallel to the surface).

\begin{figure}
  \includegraphics{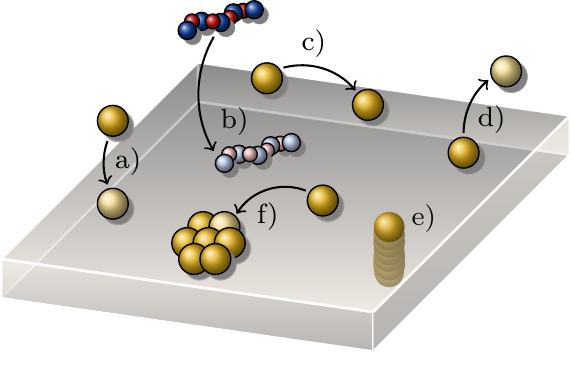}\hspace*{-0.9cm}
  \includegraphics{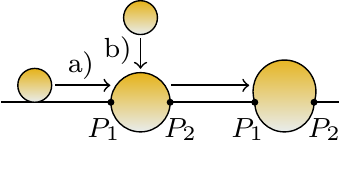}
  \caption{\textit{Left}: Sketch of the cluster processes included in the simulations: deposition of monomers\,a) and polymer\,b),
                 surface diffusion of clusters\,c), evaporation of monomers\,d), formation of metallic nanocolumns\,e),
                 cluster growth induced by surface diffusion\,f).
           \textit{Right}: Illustration of the two basic mechanisms of nanocolumn growth incorporated in the simulations.
                  column growth induced by surface diffusion\,a), column growth after direct impingement of a monomer on a cluster\,b).
                  The polymer surface moves upwards during the deposition process due to arrival of new organic molecules from the gas phase.
          }
  \label{fig:growth_mechanism}
\end{figure}

Metal atoms and clusters are modeled within the framework of the liquid drop model. They are considered as non-decomposing spheres with constant
density regardless of size, where a single atom is assumed to have a radius $r_1=0.145$\,nm. Atoms are deposited randomly on the surface and
start immediately to perform isotropically distributed surface diffusion jumps with a diffusion frequency \difffreq~ and an averaged diffusion
jump length $l$ which is chosen to be $l=0.6\,\textmd{nm}$, which is approximately the diameter of a polycarbonate chain.
Clusters\footnote{for simplicity atoms are also referred to as clusters from now on.} obey two different growth mechanism: the first one for liquid clusters is the
fusion of two clusters to a larger one according to the reaction scheme $M_n\,+M_m\rightarrow M_{n+m}$, where the subscript labels the number of
atoms the cluster consists of. Merging of clusters occures when the distance of two clusters falls below half of the jump length\,(0.3\,nm)
and is assumed to take place without any finite equilibration time, i.e. the spherical shape of the new cluster is reached immediately after
encountering of the two constituting clusters. The second growth mechanism clusters undergo leads to the formation of elongated nanocolumns growing
into the direction of the surface. It is initialized when the radius of one of two merging clusters is above the melting radius \meltrad,i.e.
when the number of atoms $N$ exceeds a critical value $N_m$. It occures when a monomer impinges directly into the interaction region of a preexisting
cluster\,(which is a circle with radius of 0.3\,nm) which is partially buried by the surface, or a cluster merges with a partially buried cluster
after surface diffusion. We assume that the new cluster does not reach a spherical shape after equilibration. Instead the incoming cluster coalesces
with the part of the buried cluster which projects beyond the surface. These growth mechanisms are subject to two boundary conditions: The first one
is volume conservation. Furthermore the points $P_1$ and $P_2$ in \fref{fig:growth_mechanism}\,(which stand for the circular intersection line of
the cluster with the surface) are assumed to stay constant. Hence the resulting initial column consists of two spherical caps separated by the
surface dividing the column into a buried part and a part projecting above the surface. We note that the intersection line of the clusters with the
surface moves upwards during the deposition process due to the arrival of new organic molecules. These two mechanisms of nanocolumn growth are depicted in
\fref{fig:growth_mechanism}. These mechanisms can repeat the same way with a free cluster and a pre-existing column where only the upper part of
the column is involved into the growth process.


To incorporate the effect of very low condensation coefficients \cond, which is known for metals on \teflon\cite{thran_condensation_1999,zaporojtchenko_determination_2000},
monomers are allowed to evaporate from the surface with a certain rate constant \evafreq~which is given in units of the surface diffusion rate \difffreq~and is adjusted
to the values of \cond~known from the experiments.
To model the deposition process of metal, atoms are randomly deposited on the surface of the substrate with a constant deposition rate \rate{m}.
A metal thickness of the diameter of a single atom corresponds to the amount of one monolayer (ML) of atoms which is defined to have a number density
of 10 atoms per $\textmd{nm}^2$. Simultaneously with the deposition of metal atoms a constant shift of the surface in $z$-direction with a certain
rate \rate{p} is applied to model the codeposition process\cite{bonitz_towards_2012}. Both rates are given in units of nm/s.

Very little is known from experiments about the size-dependence of diffusion coefficients of metal clusters on polymer surfaces. We use a power law
depending on the cluster size $n$, $\diffconst^n=n^{-\alpha}\diffconst^1$, $\alpha=1$, which is known from Molecular Dynamics simulations of cluster
diffusion on crystalline surfaces\cite{Jen99}.

In order to obtain the best accordance with the experiments we decided to treat the diffusion constant \diffconst~and the melting radius \meltrad~as free
parameters. 
\section{Simulation results}
To check the applicability of our simulation model we performed simulations over a broad range of parameters. For all results presented below,
the composites have a thickness of 100\,nm and a surface cross section of $350\,\textmd{nm}\times 350\textmd\,{nm}$. The metal deposition rate \rate{m}
~was kept to a constant value of 1.5\,nm/min. The ratio of deposition rates $\kappa:=\rate{m}/\rate{p}$ was regulated by tuning
the deposition rate \rate{p}~of the polymer substrate to the desired value. The melting radius of the clusters was treated as a free parameter.
In order to obtain good statistics, the results presented below are averaged over 20 runs of our code with constant parameter set. The
deviations were usually less than one percent so that they are not included into the figures. The focus of our investigations will lie on the
influence of the atomic evaporation\,(desorption from the surface) and the surface diffusion rate, which depends on the metal-polymer interaction
and hence on the condensation (or sticking) coefficient \cond. A low \cond~value is accompanied with a large diffusion length. 

The main effect which was observed during experiments is a dramatic increase of the volume filling factor with the ratio $\kappa$.
In contrast to the experiments, the simulations provide additional data such as the size distribution of clusters and the geometrical
measurements and the exact number of the nanocolumns, which allows for a more complete understanding of the self-organized process
of nanocolumn growth.

\begin{figure}[t!]
  \includegraphics{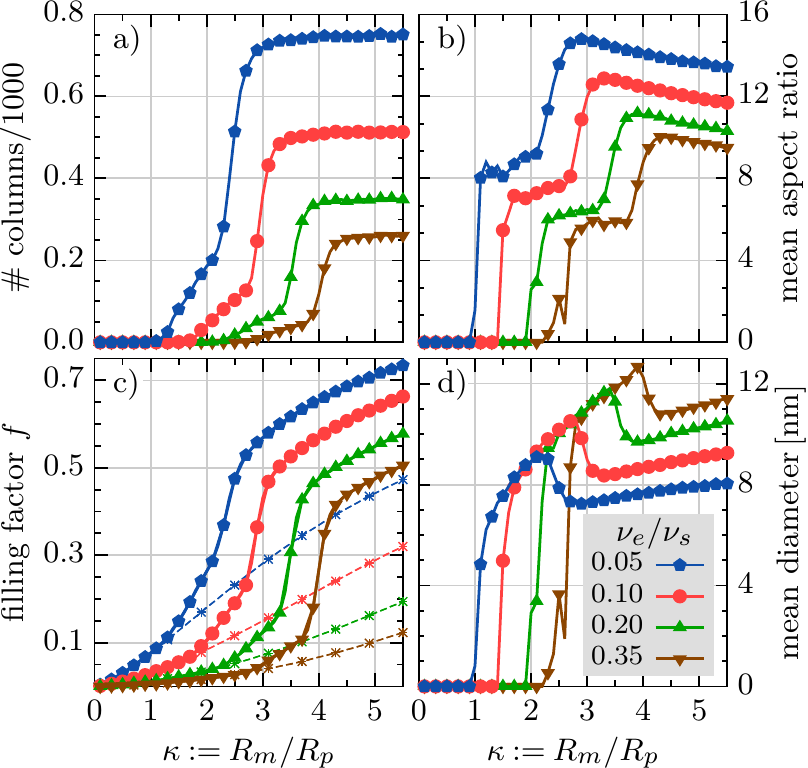}
  \caption{Simulation results of different quantities as a function of the ratio of deposition rates $\kappa$ for four different values of
           the evaporation rate \evafreq. The melting radius \meltrad~was set to 2.23\,nm and the value of surface diffusion coefficient
           \diffconst~to \expnum{1.845}{-11}\,cm\textsuperscript{$2$}/s 
           (\difffreq=\expnum{2.05}{4}\,s\textsuperscript{$-1$}).
           \textit{Upper panel}: a)\,total number of columns, b)\,mean aspect ratio (defined as the ratio of the mean column length and the mean column diameter).
           \textit{Lower panel}: c)\,metal filling factor including column growth\,(solid lines) and without column growth\,(dashed lines), d) mean column diameter.
          }
  \label{evaporation}
\end{figure}

In \fref{evaporation} simulation results of different quantities as a function of the ratio of deposition rates $\kappa$ are shown
for a system with a surface area of $350\,\textmd{nm}\times 350\textmd\,{nm}$ and a final thickness (after the termination of both deposition
processes) of 100\,nm.

The simulations clearly show a strong increase of the metal filling factor $f$ for values of $\kappa\geq 1.5$ which is related to the
formation of nanocolumns\, (cf. \fref{evaporation}\,a) and c)). The upper panel of \fref{evaporation} which shows the number of
nanocolumns indicates that there is a sharp transition from the pure spherical growth regime to a regime of column growth that coincides
with the strong increase of the filling factor. Within our model the explanation of this phenomenon is as follows:
When atoms impinge on the surface they may undergo various competing processes like surface diffusion, reemission and nucleation after
ecountering each other. One crucial point for the observed transition is the low condensation coefficient of metal on \teflon, which is
caused by the weak chemical interaction of the two components. Metal atoms\,(clusters) have to encounter each other and form nuclei that
can be stabilized in the polymer matrix and initiate the column growth. The simultaneous deposition of the polymer matrix works against
the growth of nuclei and isolates the clusters from each other. For low values of $\kappa$ the reemission of atoms and the growth of the
polymer matrix are the dominant processes and prevent the growth of clusters that are big enough to initiate the column growth.
When $\kappa$ exceeds a critical value the deposition of metal atoms plays the dominant role and the reemission and isolation of clusters
by the growing matrix is compensated by agglomeration of atoms with pre-existing clusters. As a consequence the growth of clusters is
strongly accelerated and some clusters reach the critical cluster size to initiate column growth.
\subsection{Effect of atomic desorption}
\begin{figure}
  \includegraphics{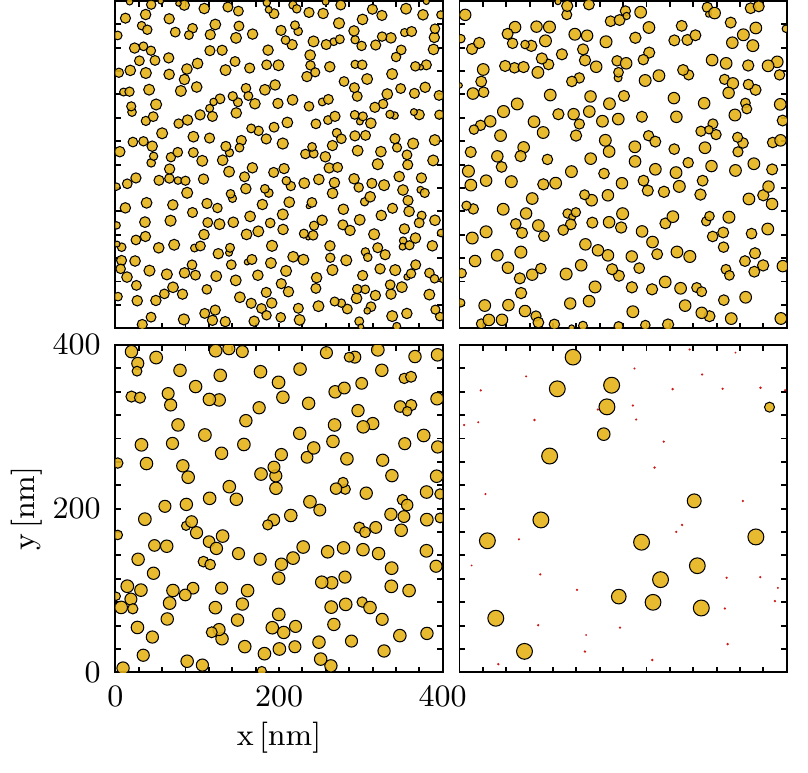}
  \caption{\label{fig:topView} Top view on the surface after termination of both deposition processes for desorption of metal atoms from the surface of the growing
           composite film with different evaporation rates and $\kappa=0.4$.
           The evaporation rates double from \expnum{2.05}{3}\,s\textsuperscript{$-1$}/s in the upper left figure to \expnum{8.2}{-1}\,s\textsuperscript{$-1$}
           the in lower right figure.
           The rest of the parameter set is the same as in \fref{evaporation}. Columns are depicted as blue circles and clusters as red ones.
          }
\end{figure}
In \fref{evaporation} the effect of the desorption with an evaporation rate \evafreq~is shown for different column geometries and the metallic filling factor.
The whole range of investigated values of $\kappa$ can be divided into four regions: The first region is characterized by pure spherical
growth and a relatively linear increase of the metal filling factor, see \fref{evaporation}\,c). In the second region the column growth is
initiated, see \fref{evaporation}\,a), and with an increasing desorption rate and the concomitant increase in the surface diffusivity, the
filling factor starts to increase nonlinearly and the transition to the columnar growth regime is shifted to larger values of
$\kappa$\,(cf. \fref{evaporation}\,a)). This effect can be easily understood in terms of the underlying column growth model: As explained in
sec.\,\ref{sec:KMC_description}, clusters have to grow beyond the melting size to act as initial nuclei for column growth. When the desorption
rate of atoms from the surface increases the growth of clusters is slowed down and, consequently, the probability of cluster growth can only be
enhanced by slowing down the embedding of clusters into the matrix, respectively by increasing the mobility of clusters which is both the effect
of increasing $\kappa$ (note that $\kappa$ is tuned via changing the polymer deposition rate \rate{p}).

A further effect of increasing the atomic desorption is a reduction in the total number of columns which is a direct consequence of the enhanced
probability of atomic reemission. This effect can also be seen in \fref{fig:topView} which shows a top view on the composite after termination
of deposition. Here one can also see how columns become thicker with higher desorption rates.
\begin{figure}[t!]
  \includegraphics{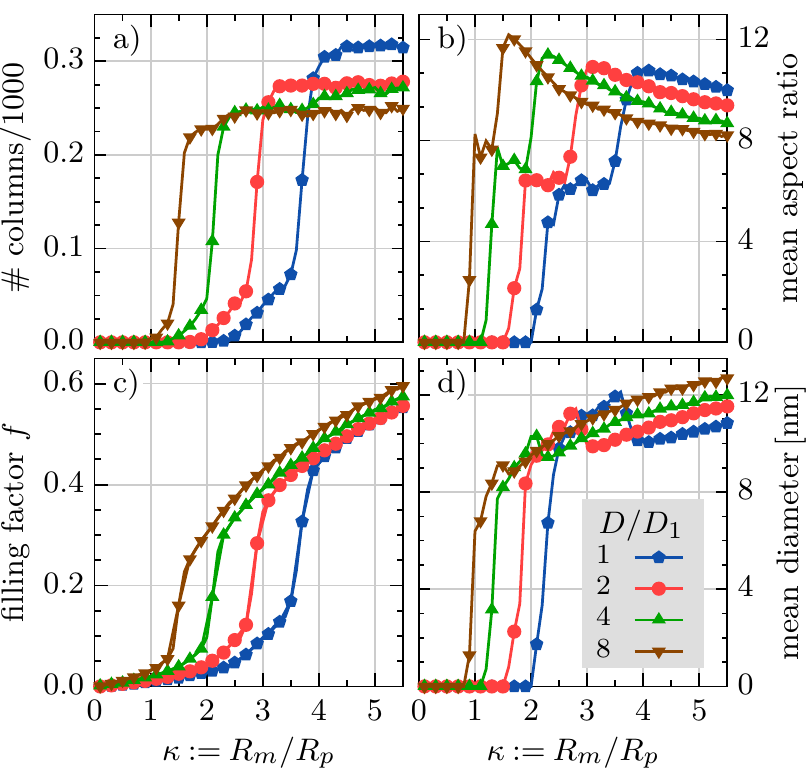}
  \caption{Same as \fref{evaporation}, but for the case of different surface diffusion coefficients \diffconst.
           The constant $\diffconst_1$ is set to \expnum{1.845}{-11}\,cm\textsuperscript{$2$}/s.
          }
  \label{fig:fill_fac_depo}
\end{figure}
The second stage is characterized by a relatively small rise of the number of columns with increasing $\kappa$. During this stage the columns
are steadily growing thicker and hold an aspect ratio in a relatively small range\,(see \fref{evaporation}\,b)) implicating that the mean column
length is also increasing with a comparable rate. Here one can see that the diameter of the columns is larger the higher the desorption rate is.

The third stage can be identified by a very abrupt increase of the total number of columns. from \fref{evaporation}\,c) one can see that the
accelerated growth of columns is accompanied by a steep rise of the filling factor, which is more pronounced the higher the desorption rate is.
The simulations have shown that this stage comes along with a considerably larger number of columns extending over the whole height of the final
composite resulting in a higher metal coverage of the surface. The mean column-diameter is decreasing during this stage 
due to the increase in the number of nuclei for column growth causing a clearly higher aspect ratio as the mean column length has already reached
its final value. The simulations have shown that the average height of the nanocolumns is some nm more than the height of
the surface of the substrate, see \fref{fig:col_length} what means that the majority of the columns now projects over the surface. Consequently
the incoming atoms are distributed to a larger number of columns as in the previous stage. 

During the fourth growth stage, whose beginning is located between $\kappa\approx3.0$\,(see the blue curve in \fref{evaporation}\,c)) and
$\kappa\approx4.2$\,(see the brown curve in \fref{evaporation}\,c)), the number of columns and their averaged length stays constant. Only
the filling factor and the diameter, respectively the aspact ratio increase due to the higher amount of metal atoms impinging on the surface.

\subsection{Influence of surface diffusion}
\begin{figure}[t!]
  \includegraphics{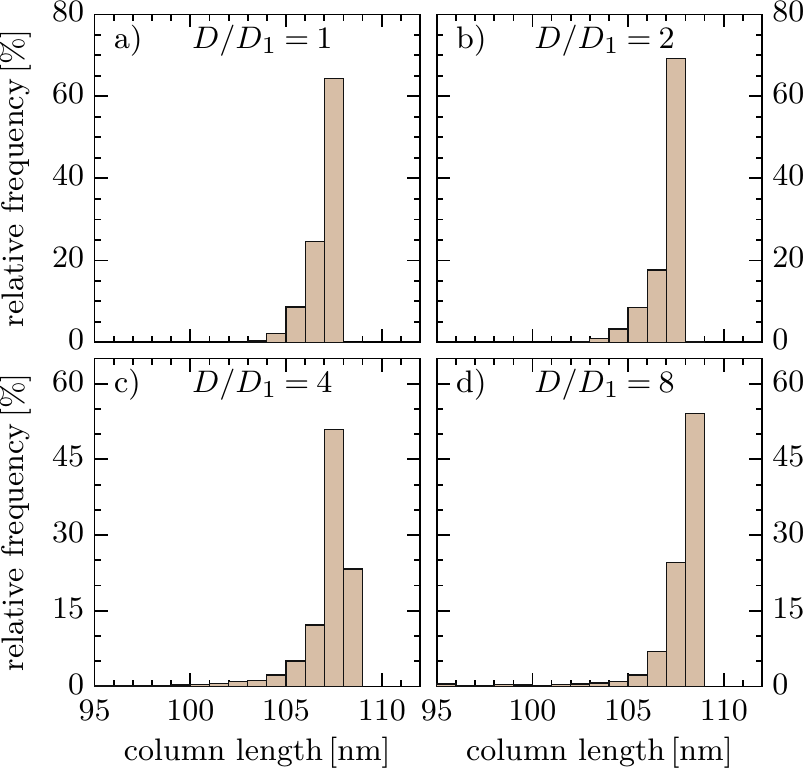}
  \caption{Distribution of column lengths for $\kappa=4.0$ depicted in a histogram with a resolution of 1\,nm. The parameter set is the same
           as in \fref{fig:fill_fac_depo}.
          }  
  \label{fig:col_length}
\end{figure}
\fref{fig:fill_fac_depo} shows the same quantities as \fref{evaporation} but for different surface diffusion coefficients \diffconst~and
constant desorption rate \rate{e}. The main effect resulting from increasing the surface diffusion coefficient is a shift of the transition
from spherical growth to columnar growth to smaller values of $\kappa$\,(see \fref{fig:fill_fac_depo}\,a) and c)). The reason for this shift
is obviously the accelerated growth of clusters caused by their higher mobility: Whilst for slow surface diffusion the cluster growth by
direct impingement of atoms on preexisting clusters plays an important role, for fast surface diffusion the probability of clusters
encountering each other during surface diffusion is considerably enhanced. As a consequence clusters reach the critical nucleus size to
induce the columnar growth already for smaller values of $\kappa$. Interestingly, the filling factors for higher values of $\kappa$\,
($\geq 4.4$) lie very close together. Our simulations have not shown a tendency of the surface diffusion coefficient to influence the filling
factor.

The diffusion constant has also no major effect on the other quantities after the transition to columnar growth. \fref{fig:fill_fac_depo}\,a)
shows that for all simulations the number of columns lies between 200 and 300 what corresponds to a surface density of columns
between $1.63\mspace{-1mu}\times\mspace{-1mu}10^{-3}$\,nm\textsuperscript{$-2$} $2.12\mspace{-1mu}\times\mspace{-1mu}10^{-3}$\,nm\textsuperscript{$-2$}
where the density is higher the lower is the surface diffusion coefficient. As one can see in \fref{fig:fill_fac_depo}\,d) the thickness of the
columns follows the opposite trend: A higher surface diffusivity leads to thicker columns, which is not surprisingly as clusters grow faster.
The aspect ratio shows a different behavior, cf. \fref{fig:fill_fac_depo}\,b). This implies that, in the simulated range of \diffconst, the
averaged length of the nanoclumns stays nearly constant.

\fref{fig:col_length} shows the distribution of column lengths after termination of both deposition processes for different surface
diffusion coefficients. All columns project some nm beyond the surface\,(which lies at 100\,nm) and that the
distribution of lengths is very narrow with a maximum at about 108\,nm for all values of \diffconst.

\section{Conclusion}
In conclusion, we presented new experimental data on the formation of magnetic Fe-Co-Ni nanocolums in a Teflon AF matrix upon co-evaporation
of the metallic and organic components, extending the temperature range of deposition to low temperatures and reporting also magnetic properties.
The nanocolums contain domain walls perpendicular to the column axis which is also the easy axis of magnetization. The formation of the nanocolums
was modeled in terms of Kinetic Monte Carlo simulations which provide an even quantitative agreement with the key experimental observations such
as the behavior of filling factor and column diameter. Unlike earlier approaches, the present simulations are able to explain the transition from
growth of spherical clusters to nanocolumn formation. It was shown that, in addition to a low metal condensation coefficient on the organic surface
and a high deposition rate ratio of metallic vs organic components, the solidification of the spherical nanoparticles at a critical radius has to
be taken into account.

With respect to applications, the results show how to tailor the nanocolumnar structure in the matrix and which parameter ranges are accessible.
The implications are not restricted to organic matrices but should hold for designing functional metal-dielectric nanocomposites in general, particularly
with respect to magnetic and plasmonic applications. The magnetization of the nanocolumns can be tuned, for instance, via their length and separation.
Moreover, an electrical field can be applied during deposition to orient the easy axis of magnetization. In particular, composites containing oriented
nanoparticles with a small aspect ratio are very interesting for high-frequency magnetic materials up to the GHz range\cite{ramprasad_magnetic_2004}.

\begin{acknowledgments}
This work was supported by the German Research Foundation (DFG) within the framework of the Collaborative Research Center SFB Transregio 24, 
projects A5, A7 and B13. The authors would like to thank Stefan Rehders for setting up the evaporation chamber and his continuous technical support.
The magnetization measurements were performed by Michael Frommberger in the group of Eckhard Quandt.
\end{acknowledgments}

\bibliographystyle{apsrev4-1}

%

\end{document}